\documentclass[conference]{IEEEtran}
   \usepackage[pdftex]{graphicx}
   \graphicspath{figs}
   \DeclareGraphicsExtensions{.pdf,.jpeg,.png}
\usepackage{multirow}
\usepackage{cite}
\usepackage[cmex10]{amsmath}
\usepackage{tablefootnote}
\begin{document}
\title{Selective Decoding in Associative Memories Based on Sparse-Clustered Networks}
\author{\authorblockN{Hooman Jarollahi, Naoya Onizawa, and Warren J. Gross}
\authorblockA{Department of Electrical and Computer Engineering, McGill University, Montreal, Quebec H3A 0E9}
Email: \{hooman.jarollahi, naoya.onizawa\}@mail.mcgill.ca, warren.gross@mcgill.ca}
\maketitle

\begin{abstract}
Associative memories are structures that can retrieve previously stored information given a partial input pattern instead of an explicit address as in indexed memories. A few hardware approaches have recently been introduced for a new family of associative memories based on Sparse-Clustered Networks (SCN) that show attractive features. 
These architectures are suitable for implementations with low retrieval latency, but are limited to small networks that store a few hundred data entries. In this paper, a new hardware architecture of SCNs is proposed that features a new data-storage technique as well as a method we refer to as Selective Decoding (SD-SCN). The SD-SCN has been implemented using a similar FPGA used in the previous efforts and achieves two orders of magnitude higher capacity, with no error-performance penalty but with the cost of few extra clock cycles per data access. 
\end{abstract}
\IEEEpeerreviewmaketitle
\section{Introduction}
As opposed to indexed memories, Associative Memories (AMs) 
retrieve data given a partial input pattern. AMs are attractive in applications pertaining to data mining and implementation of sets such as multiple-field search-engines \cite{XML2010}, where data is accessed by an explicit address that needs to be generated after often a large number of look-up operations. 
%

The state-of-the-art Hopefield Neural Network (HNN) is a classical AM introduced in \cite{Hopfield1982}. When implemented in hardware, HNNs can quickly retrieve partial input patterns in small networks capable of storing only a few data entries (messages) \cite{Gripon2011b}. However, HNNs suffer from major drawbacks: First, in order to increase the size of the memory, the length of the messages need to be unnecessarily increased. Thus, due to the limited availability of the physical memory, the number of different messages it can store (diversity) is confined. Second, the ratio between the number of information bits that it can store to the memory bits that it requires to store them (efficiency) approaches zero as the memory size is increased \cite{Gripon2011b}. 

Recently, a new class of AMs has been proposed by \cite{Gripon2011b, Gripon2012}, in which the authors address the drawbacks of HNNs. These AMs are based on Sparse-Clustered Networks (SCNs), and can achieve significantly larger diversities and efficiencies compared to those of HNNs \cite{Gripon2012}. Furthermore, the hardware implementation of SCNs has a lower level of complexity compared to that of HNNs since the value of the nodes (neurons) and the edges (links) are binary. A proof-of-concept hardware architecture for SCNs was first proposed in \cite{Jarollahi2012}, and achieved a significant speed-up of $\approx2000\times$ compared to that of its CPU-based counterpart. However, the architecture included resource-hungry max-functions and large-input adders that limited the scalability to few dozens of messages. A reduced-complexity algorithm and architecture were later proposed in \cite{ICASSP2013}, where the authors eliminated the max-function and the adders by combining information from multiple ambiguous neurons in a cluster instead of treating them independently. However, the previous works in this area still employ massively-parallel logic-gates, and independently-accessed on-chip registers that limit the scalability of such implementations to store a few hundred messages. 

In this paper, we propose a hardware architecture that is based on a new data-storage technique and a decoding structure we refer to as Selective Decoding (SD). The SD-SCN can achieve a capacity (number of stored data bits) that is two orders of magnitude larger compared to that of the previous efforts using a similar FPGA. Furthermore, it retrieves data with no error-performance penalty although it can cost a few extra clock cycles per message retrieval operation. The algorithm behind SCN is briefly explained from a hardware design perspective in Section \ref{sec:GBNN}, and the proposed hardware architecture is introduced in Section \ref{sec:Prop_Arch}. Section \ref{sec:res} summarizes the results followed by conclusions in Section \ref{sec:concl}.
\vspace{-1 mm}
\section{Selective Decoding in Sparse-Clustered Networks}
\label{sec:GBNN}
\begin{figure}[!t]
\centering
\vspace{-4 mm}
\includegraphics[width=3.2in]{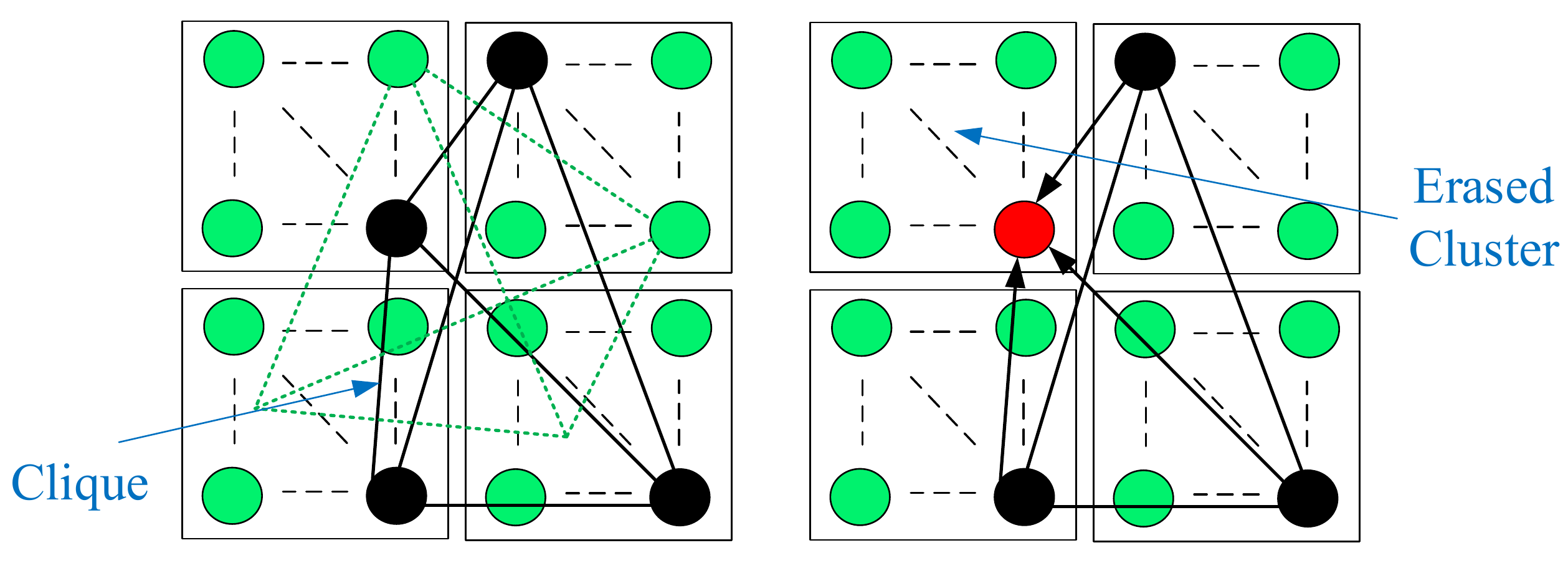} 
\vspace{-3 mm} 
\caption{Representation of neurons, clusters and stored cliques in the message storage process (left), and a neuron in an erased-cluster being activated during the global decoding process of the message retrieval (right).}
\vspace{-5 mm}
\label{fig_fanals}
\end{figure}

The recent work on SCN-based AMs \cite{Gripon2011b} has been mainly inspired from HNNs \cite{Hopfield1982}. As shown in Fig. \ref{fig_fanals}, these types of AMs are made out of $n$ neurons that are arranged into $c$ equally-sized clusters and receive or propagate information by their binary values. The connections between these neurons are also binary indicating the existence of a connection. Each cluster is associated with a part of an input pattern that is also equally divided into $c$ sub-patterns. Furthermore, the network is $c$-partite, which means that two neurons in a same cluster cannot be connected. A partial input pattern is presented and a sparse set of neurons are activated which represent the matching learned pattern. The set is then encoded to form the full output pattern, also called a \emph{clique} \cite{Gripon2011b}.

%
\subsection{Message Storage}
\label{sec:GBNN_training}
In order to store data into the memory, a message $m$ of $K$ bits is first divided into equal parts of $\kappa$-bits each resulting in the creation of $c = K/ \kappa$ sub-messages. The network has a distinct neuron dedicated to each sub-message and thus, there exists $l=2^\kappa$ binary neurons. Each sub-message is simply mapped into a neuron in its corresponding cluster using a direct conversion of its binary value to an integer number representing the index of the neuron it is being mapped into. Once all sub-messages have been mapped, all of the corresponding mapped neurons are connected to construct a fully-connected pattern known as a $clique$ \cite{Gripon2011b} as shown in Fig. \ref{fig_fanals}. The connections in a clique are stored in a memory module that is accessed during the retrieval process.  

In this paper, $s(n_{(i,j)})$, $v(n_{(i,j)})$ and $w_{(i,j)(i',j')}$ refer to the integer score value, the binary value of the $j$-th neuron in the $i$-th cluster, and the binary interconnection value between the $j$-th neuron of the $i$-th cluster to the $j'$-th neuron of the $i'$-th cluster respectively.  




%
\subsection{Message Retrieval}
\label{sec:GBNN_retrieval}
To retrieve a message, we first determine which neurons should be activated given a partial or erroneous input pattern and using a similar method as in the message storage. This process is called Local Decoding (LD). Since bit-erasers may also exist in a sub-message, multiple neurons for a cluster might be activated. We refer to these cases where multiple neurons are activated as \emph{ambiguities}. If a sub-message is entirely erased, all of the neurons are activated in its corresponding cluster. We then use the interconnection weights to determine which neurons should remain activated based on a previously learned pattern as shown in Fig \ref{fig_fanals}. This process, also called Global Decoding (GD), iterates until only one neuron per cluster is activated or the number of activated neurons is not changed. The active set of neurons is then encoded to form the output. If no recovery is possible, an \emph{error} has occurred.
%
\subsubsection{Local Decoding}
In the LD process, each input sub-message is mapped into an index of a neuron in its corresponding cluster.  In this process the value of a neuron in other clusters does not affect that of one in the decoding cluster, hence the name. 
The conventional mapping process as described and implemented in \cite{Gripon2011b, Gripon2012, Jarollahi2012} employs matrix multiplication to compute a score for each neuron followed by finding the maximum score and the winner-take-all rule. 

%
The complexity of the last step can be reduced as shown in \cite{ICASSP2013}, where the authors demonstrated that the max-function can be eliminated by realizing that the maximum score of a neuron can be pre-determined without using matrix multiplication and the use of resource-hungry max-functions but instead, based on the number of erased bits (erasures). The maximum score can be computed by subtracting the number of erased bits, $n_e$ from $\kappa$ as follows:   
\vspace{-1 mm}
\begin{equation}
v(n_{(i,j)}) \leftarrow \left\lbrace \begin{array}{cc}
                                1 , & \mbox {if } s(n_{(i,j)}) = \kappa - n_e \\
                                0 , & \mbox {otherwise}
                               \end{array} \right.
\label{equ_ld1c}
\end{equation}
Therefore, in applications where there could only exist erased clusters rather than erased bits, the score is either equal to zero or $\kappa$. Therefore, the local decoding complexity can be reduced to simply converting the sub-messages to their integer values and activating the neurons with these integer indices. An erase-flag can be externally generated to enable the activation of all of the neurons in the erased cluster during the first iteration. In this paper, we focus on hardware implantation of the last scenario.

\subsubsection{Global Decoding}
In order to remove the ambiguous neurons after LD, an iterative process is performed using all of the neurons in the clusters other than the one in which a neuron is being decoded. After the first iteration, it is possible that more than one neuron is still activated in the erased (or partially-erased) cluster. The iterations can continue until the results converge, i.e. no more ambiguities remain. Finally, the index of the activated neuron in each cluster is encoded to form the erased sub-message. 

It has been be shown in \cite{Gripon2012, ICASSP2013} that during the GD process, the value of a neuron can be computed to become or stay activated (if previously activated by the LD) if and only if it receives at least one signal from every other cluster than itself. In other words, the ambiguities of other clusters will not have an effect on the value of the neuron being computed in GD. As shown in \cite{ICASSP2013}, the GD process reported in \cite{Gripon2012} can be reformed as:
\vspace{-3 mm}
\begin{equation}
v(n_{(i,j)}) = \bigg( \bigwedge_{j'=1}^{c-1}  \bigvee_{i'=1}^l \left( w_{(i,j)(i',j')}v(n_{(i',j')})\right)\bigg) \bigwedge v(n_{(i,j)} )                             
\label{equ_gd_ICASSP}
\end{equation}
%

\noindent where $\bigvee_{i'=1}^l$ performs an $l$-input OR function, and $\bigwedge_{j'=1}^{c-1}$ performs a ($c-1$)-input AND function. This algorithm is suitable for scenarios when bit erasures exist, and will not operate for erroneous input corrections. 

The conventional GD architectures \cite{Jarollahi2012, ICASSP2013} are based on a technique we refer to as Massively-Parallel Decoding (MPD). In MPD-based hardware architectures, there exists massively-parallel logic gates and independently-accessed on-chip registers that perform the $w_{(i,j)(i',j')}v(n_{(i',j')})$ operations in (\ref{equ_gd_ICASSP}) for all neurons in parallel and limiting the scalability. However, we can simplify these operations given that $v(n_{(i',j')})$ is pre-determined by the LD before the iterations or the GD after the first iteration as follows: 
\vspace{-1 mm}
\begin{equation}
v(n_{(i,j)}) = \bigg(\bigwedge_{j'=1}^{c-1}\bigvee_{\substack{i'=1\\v(n_{(i',j')})=1}}^l \left(w_{(i,j)(i',j')}\right)\bigg) \bigwedge v(n_{(i,j)}). 
\label{equ_gd_prop}
\end{equation}
%
Therefore, we can rearrange the conventional GD algorithm shown in (\ref{equ_gd_ICASSP}) in favour of the hardware scalability by adding a condition that will not affect the error performance. Furthermore, we have observed in simulation results that after the first iteration, the maximum number of activated neurons per cluster among all clusters is small. This number, we refer to as $\beta$, is also the number of serial accesses to the RAM blocks in an FPGA to retrieve $w_{(i,j)(i',j')}$ values. Therefore, the serialization of the GD process to read the links will cost only of few extra clock cycles compared to its fully-parallel counterpart.

\vspace{-3 mm}
\section{Proposed Architecture}
\label{sec:Prop_Arch}
The hardware architecture of SD-SCN presented in this section is based on (\ref{equ_gd_prop}). First, we will show how the links between the neurons are stored in the Link Storage Module (LSM) during the write process. Then, we will present the architectures and the function of the constituent blocks including the LD, the GD, and the Serial Pass Module (SPM).
\vspace{-2 mm}
\subsection{Link Storage Module}
\begin{figure}[!t]
\centering
\vspace{-4 mm}
\includegraphics[width=3.0in]{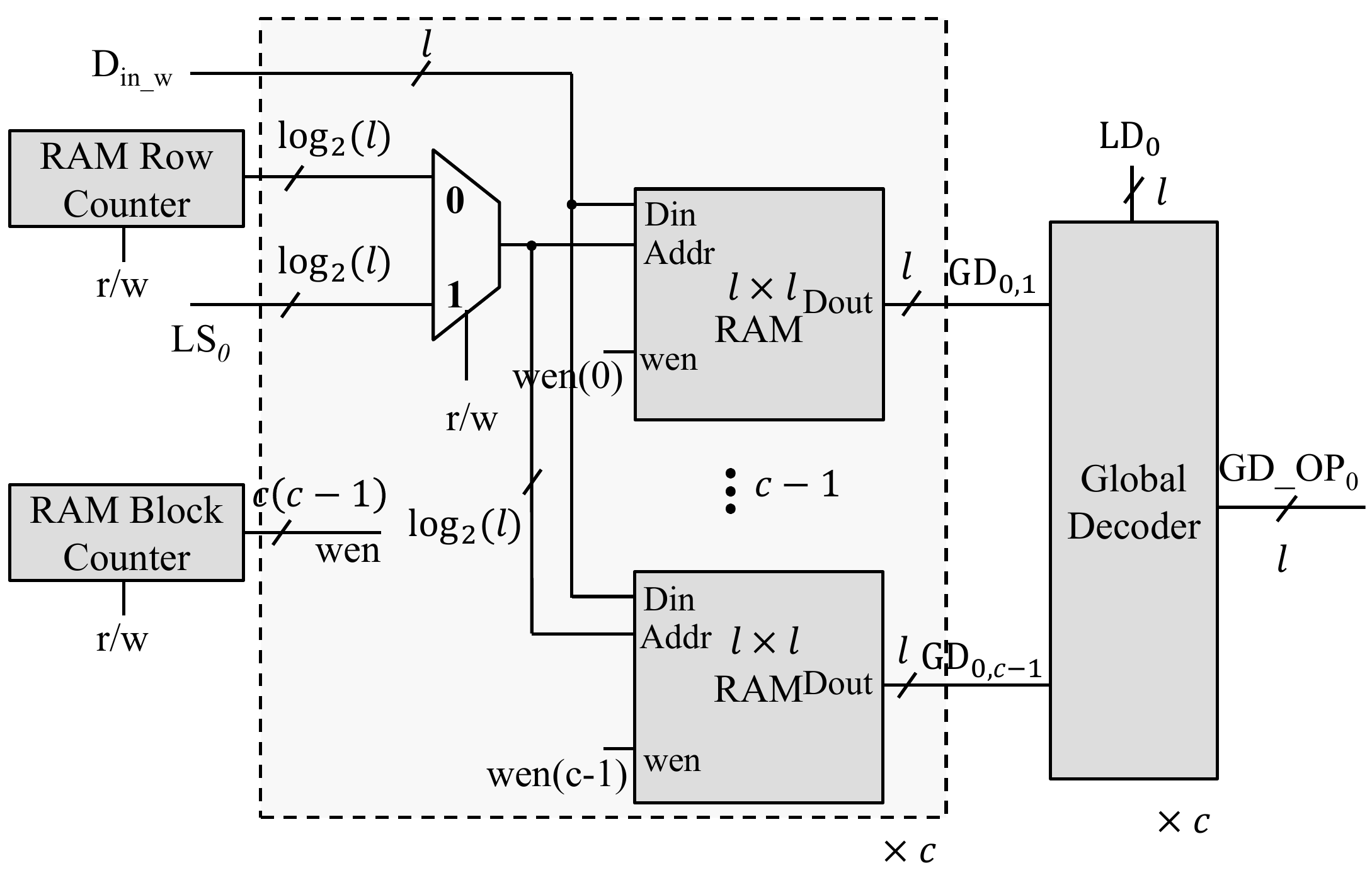} 
\vspace{-3 mm} 
\caption{Architecture of the LSM.}
\label{fig_RAM}
\end{figure}

\vspace{-1 mm}
The memory architectures in \cite{Jarollahi2012} and \cite{ICASSP2013} are similar, and both employ massively-parallel on-chip registers (flip-flops) that are accessed simultaneously in the GD. Therefore, scaling these architectures are challenging due to the drastic growth of the number of interconnections from these registers to the logic elements as the network grows, aside from the limitation in the number of available registers in an FPGA. Following (\ref{equ_gd_prop}) in SD-SCN, we will show how to replace these registers with scalable RAM blocks. As shown in Fig. \ref{fig_RAM} the LSM includes $c(c-1)$ RAM blocks each arranged into $l \times l$ bits. Each RAM block stores the links between the neurons between two clusters. A matrix of size $c(c-1)l \times l$ containing the links for all messages can be first generated by the CPU that communicates the user's inputs with the SD-SCN, or on-chip as in \cite{Jarollahi2012, ICASSP2013}. During the write-mode (indicated by the r/w signal), the links are transferred into the RAM blocks row by row. If $l$ is larger than the number of pins on the FPGA, this process can be serialized by dividing $l$ bits into multiple parts and transferring each part sequentially. Once the read/write signal (r/w) is in the write-mode, the RAM row counter starts counting from $0$ to $\text{log}_2(l)-1$ providing the address to the RAM block in which the data is being written. Each RAM block is enabled for writing using the RAM block counter that sequentially selects the memory blocks during the write mode. 

During the message retrieval process (read-mode), the LSM is accessed using $c(c-1)$ inputs that are selected from either the output of the LD in if its corresponding cluster is not erased, or from the outputs of the GD after each iteration after serially processed by the GD Serial-Pass Module (SPM). In case of a cluster erasure, the access to LSM is skipped for that particular cluster and the output of the LD is directly passed to the GD.
\vspace{-4 mm}
\subsection{Local Decoder (LD)}
\begin{figure}[!t]
\centering
\vspace{-4 mm}
\includegraphics[width=3.0in]{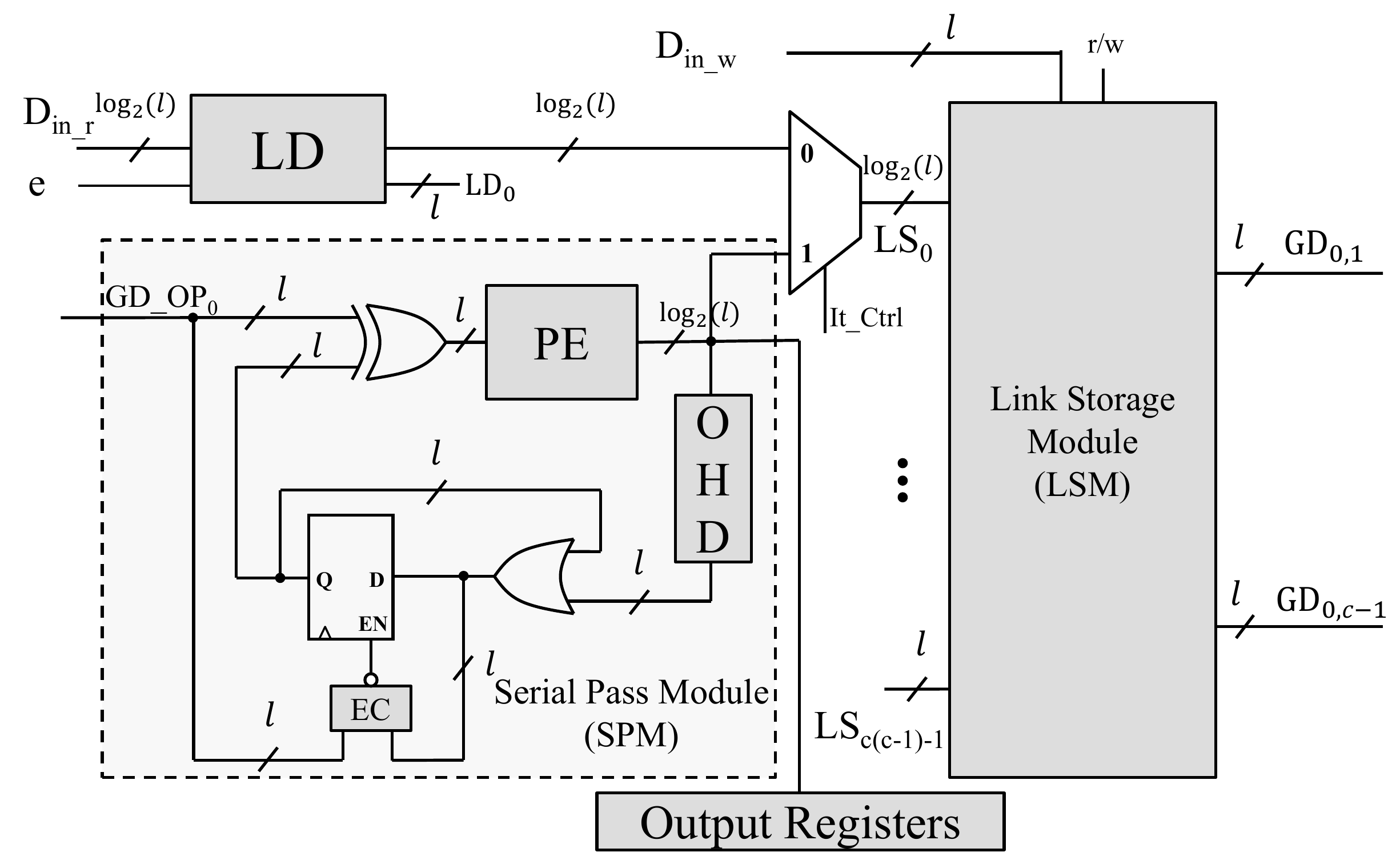} 
\vspace{-3 mm} 
\caption{Integration of the SPM with the LD and the LSM.}
\vspace{-5 mm}
\label{fig_sp}
\end{figure}

As shown in Fig. \ref{fig_sp}, the LD outputs two signals: 1) a direct connection between a log$_2(l)$-bit vector from the input if the cluster is not erased and 2) an $l$-bit vector generated by a One-Hot-Decoder (OHD) that is used in the GD. If a cluster is erased, as indicated by an erase flag, $e$, the LD outputs a vector containing all ones.
\vspace{-3 mm}
\subsection{Global Decoder (GD)}
\begin{figure}[!t]
\centering
\vspace{-2 mm}
\includegraphics[width=2.7in]{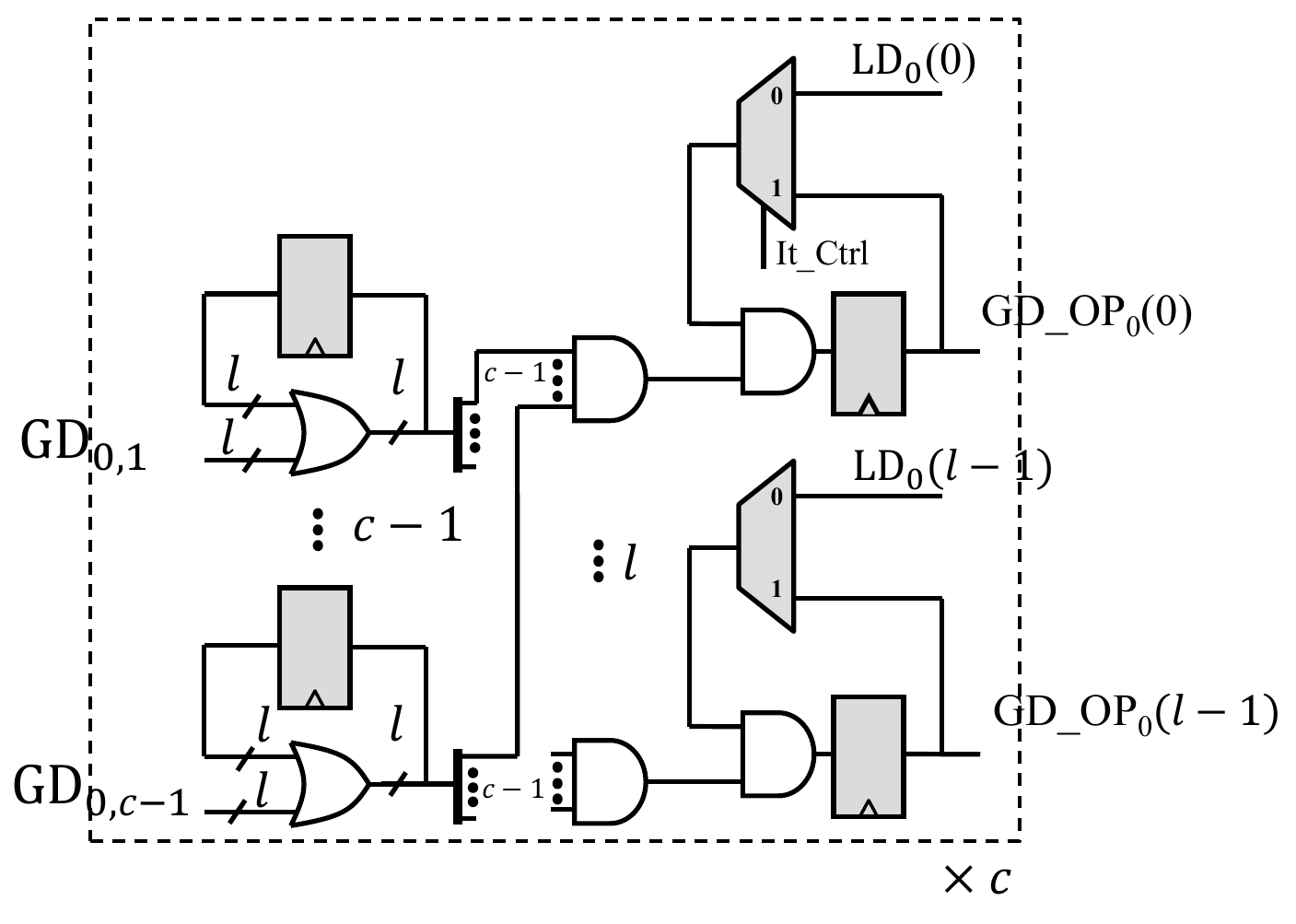} 
\vspace{-3 mm} 
\caption{Schematic of the global decoder based on the selective-decoding technique.}
\vspace{-5 mm}
\label{fig_SD}
\end{figure}

The architecture of the GD that is presented in \cite{ICASSP2013} is based on MPD algorithm shown in (\ref{equ_gd_ICASSP}). This decoding rule performs the $w_{(i,j)(i',j')}v(n_{(i',j')})$ operations in (\ref{equ_gd_ICASSP}) using $c(c-1)l^2$ two-input AND gates on the links and the values of the neurons regardless of the output values obtained from the LD or the GD after each iteration in spite of the fact that a zero value of the neuron will automatically results in a zero in the corresponding AND operation. Therefore, the hardware complexity is higher than the scenario, where instead of performing the AND operation, we simply examine whether a connection exists or not given an activated neuron.  The proposed GD is based on the SD algorithm (\ref{equ_gd_prop}) explained earlier.  The SD algorithm reduces the hardware complexity as no additional computations would be required for those neurons that are not previously activated. A simplified schematic view of the GD is depicted in Fig. \ref{fig_SD}, where the inputs are the binary link values ($w_{(i,j)(i',j')}$) retrieved from the RAM blocks of the LSM. Since multiple neurons might be activated after each iteration, and that the input address of each RAM block should be a single address, the outputs of the GD are connected to the SPM that sequentially passes the outputs of the GD to the RAM blocks consuming a few clock cycles. 

%
%
\begin{table*}[t]
\renewcommand{\arraystretch}{0.8}
\begin{center}
\caption{Result Comparison ($\beta=2$, $it=4$, $c=8$, DNF: Does Not Fit).}
\vspace{0mm}
\begin{tabular}{c||c|c|c|c|c|c|c}
\hline
&   \multicolumn{2}{c|}{ISCAS 2012 \cite{Jarollahi2012}}  & \multicolumn{2}{c|}{ICASSP 2013 \cite{ICASSP2013}} & \multicolumn{3}{c}{Proposed}\\
\hline
\hline
Number of Messages (M)    & 64   & 1018 &  64    & 1018  &  64   & 1018  & 39,754   \\
\hline
Number of Neurons (n)     & 128  & 512  &  128   & 512   &  128  & 512   & 3200     \\
\hline
Capacity (Kbits)  & 2.05 & 48.86&  2.05  & 48.86 & 2.05  & 48.86 & 2862.29   \\
\hline
LUTs              & 35,224 & DNF& 12,341 & DNF   & 1,956 & 8,082 & 47,352    \\
\hline
Registers         & 15,783 & DNF&  15,035& DNF   & 1,333 & 5,239 & 32,609     \\
\hline
BRAM Bits         & 0      & 0  &  0     & 0     & 14,336 & 229,376 & 8,960,000 \\
\hline
Max. Freq. (MHz)  & 107.15 & DNF  &  205.21    & DNF  &  261.85   & 159.29  & 72.04  \\
\hline
Access Delay (clock cycles)   & \multicolumn{4}{c|}{$1+it$} &  \multicolumn{3}{c}{$2+(\beta+1)\times (it-1)$}      \\
\hline
\end{tabular}
\vspace{-5mm}
\label{table_res}
\end{center}
\end{table*}

The number of clock cycles that the SPM requires to address the GD outputs depends on the maximum number of activated neurons among all clusters after the first iteration. This value, $\beta$, was simulated in software with respect to the density for two networks both consisting of 8 clusters ($c=8$), one with 128 and the other 3200 neurons. The networks were loaded using uniformly-random messages. $\beta$ was measured using 1000 random inputs with 50\% erased clusters. For a reference density (0.22 as suggested in \cite{Gripon2011b}), $\beta$ is equal to two. If the data is not uniformly random or other densities are needed, $\beta$ can be adjusted accordingly.

In order to implement the $\bigvee_{i'=1}^l$ operations in (\ref{equ_gd_prop}), the outputs of the SPM are accumulated using two-input OR gates and a feedback register over $\beta$ clock cycles. Then, $\bigwedge_{j'=1}^{c-1}$ operations are performed on the outputs using ($c-1$)-input AND gates. Finally in order to implement the AND-operation in the last part of (\ref{equ_gd_prop}), also known as the \emph{memory effect} \cite{Gripon2011b}, a two-input AND gate is considered and its output is registered. One of the inputs of the AND gate is the output of a multiplexer that controls data-flow from the LD or the output of the GD. During the first iteration, the output of the multiplexer is set to be the output of the LD (using \emph{It\_Ctrl} signal) since no output is yet computed by the GD.
\vspace{-2 mm}
\subsection{Serial-Pass Module}
Because the RAM structure does not allow simultaneous accesses, and that the GD can produce more than one activated neuron in an erased cluster especially after the first iteration, the $l$-bit values in each cluster are connected to a module to sequentially process the activated neurons. As shown in Fig. \ref{fig_sp}, the SPM uses a priority encoder (PE) that  prioritizes the activated neurons in each cluster from the most significant bit to the least within a deterministic number of clock cycles ($\beta$). The width of the PE's output is log$_2(l)$-bits which is equal to address width of the RAM blocks in the LSM. The role of the parallel two-input XOR gates that are connected before the PE is to convert the values of a recently processed inputs to zero after a clock cycle so that the PE can prioritize all of the activated neurons. The parallel two-input OR gates and the flip-flops perform the inverse operation of serializing i.e. accumulating previous outputs such that zeroing all previously operated input bits using the XOR gates becomes possible by remembering the previously processed data. An equality comparator (EC) generates the enable signals for the registers.  A One-Hot Decoder (OHD) is used to convert log$_2(l)$-bits back to $l$ bits so that the inverse serialization process can be possible. After the iterations complete, the final output is registered from the output of the PE of each cluster. 
\vspace{-1 mm}
\section{Results}
\label{sec:res} 
The SD-SCN has been implemented and verified using Altera Stratix IV (EP4SGX230KF40C2) FPGA as used in \cite{Jarollahi2012, ICASSP2013}. As illustrated in Table \ref{table_res}, the previous SCN architectures can not scale to a network with 512 neurons or larger whereas SD-SCN can contain 3200 neurons that can store $621\times$ as many messages compared to that of the previous works at a reference density of 0.22. The value of $\beta$ is simulated to be equal to two for this network, and with $it=4$ for the number of iterations, the network can converge to the final output status.
\vspace{-2 mm}
\section{Conclusion}
\label{sec:concl} 
In this paper, a new hardware architecture is proposed for associative memories based on sparse-clustered networks. Previous efforts in this area employed massively-parallel logic gates, interconnections, and on-chip registers that limited the scalability of the memory. The proposed architecture (SD-SCN) employs new message-decoding and data-storage techniques in hardware that remove these gates and registers, and as a result increase the maximum number of stored data bits (using the same FPGA as the previous efforts) by two orders of magnitude. Compared to the previous architectures, SD-SCN results in no error-performance penalty, but costs of a few extra clock cycles per message retrieval operation that can be adjusted depending on the data distribution. 
\vspace{-1mm}
\bibliographystyle{IEEEtran}
\bibliography{hooman}{}
\end{document}